# Superperiods in interference of e/3 Laughlin quasiparticles encircling filling 2/5 fractional quantum Hall island

Ping V. Lin[1], F. E. Camino[2], and V. J. Goldman[1]

[1] *Department of Physics, Stony Brook University, Stony Brook, NY 11794-3800, USA*
[2] *Center for Functional Nanomaterials, Brookhaven National Laboratory, Upton, NY 11973, USA*

We report experiments in a large, 2.5 μm diameter Fabry-Perot quantum Hall interferometer with two tunneling constrictions. Interference fringes are observed as conductance oscillations as a function of applied magnetic field (the Aharonov-Bohm flux through the electron island) or a global backgate voltage (electronic charge in the island). Depletion is such that in the fractional quantum Hall regime, filling 1/3 current-carrying chiral edge channels pass through constrictions when the island filling is 2/5. The interferometer device is calibrated with fermionic electrons in the integer quantum Hall regime. In the fractional regime, we observe magnetic flux and charge periods $5h/e$ and $2e$, respectively, corresponding to creation of ten $e/5$ Laughlin quasiparticles in the island. These results agree with our prior report of the superperiods in a much smaller interferometer device. The observed experimental periods are interpreted as imposed by anyonic statistical interaction of fractionally-charged quasiparticles.

## I. INTRODUCTION

A system of electrons constrained to move in two dimensions (2D) in a strong magnetic field exhibits exact quantization of Hall conductance at certain integer and fractional Landau level fillings.[1-5] While the integer quantum Hall effect can be understood as a consequence of Landau quantization of non-interacting electrons, the fractional quantization is understood as resulting from condensation of interacting electrons into a highly-correlated incompressible fluid. The elementary charged excitations of a fractional quantum Hall (FQH) condensate are Laughlin quasiparticles possessing bizarre properties: they have fractional electric charge[3-8] and obey anyonic (fractional) exchange statistics,[9-14] intermediate between the familiar Bose and Fermi statistics.

Upon exchange of two anyons, the quantum state of the system acquires a phase which is neither 0 nor $\pi$, but can be *any* value.[15] In two dimensions, one particle adiabatically encircling another is equivalent to their exchange done twice (exchange operation squared).[9] This topologically robust property can be used to detect anyons in interference experiments, because when either bosons or fermions encircle other particles, the system's wave function acquires an integer multiple of $2\pi$ phase difference, which does not affect the interference pattern. For anyons, the acquired phase difference is, in general, non-trivial, and thus does affect the interference. This nonlocal, topological interaction of anyons has lead to several proposals to use braiding of anyons in 2D systems for topological quantum computation.[16,17]

Specifically, for charge $e/3$ quasiparticles of the filling $f = 1/3$ FQH fluid, an explicit calculation shows that the system's wave function acquires an anyonic Berry phase contribution when one Laughlin quasihole adiabatically encircles another.[11] Experiments on quantum antidots[18] and Fabry-Perot quantum Hall interferometers[14] reported Aharonov-Bohm flux period $\Delta_\Phi = h/e$ for the $e/3$ quasiparticles, while for fermionic or bosonic $e/3$ quasiparticles the expected flux period would be $h/(e/3) = 3h/e$. These experimental results were interpreted as evidence that the quasiparticles of the $f = 1/3$ FQH fluid are indeed anyons, the "missing" $4\pi/3$ phase difference supplied by the statistical Berry phase contribution, in agreement with the theory of Ref. 11. Experiments[19-26] on two-constriction electron Fabry-Perot interferometer devices in the integer quantum Hall regime, and a chiral Luttinger liquid theory[27] of such devices in the primary Laughlin states were also reported.

Less clear theoretically is the situation when different kinds of quasiparticles are involved, even for the next simplest case of the $e/3$ and the charge $e/5$ quasiparticles of the 2/5 FQH fluid, which is the simplest hierarchical "daughter state" of the 1/3 fluid.[28,12] Earlier, we reported experiments on an interferometer where $e/3$ quasiparticles of the 1/3 FQH fluid encircle an island of the 2/5 fluid.[13,29-31] The interference conductance oscillations occur as a function of magnetic field, or the island electronic charge varied by a backgate. The flux and charge periods were obtained using the Aharonov-Bohm interference area,[32,33] which, in turn, was determined either from modeling of the island electron density profile,[13] or experimentally, via scaling the Aharonov-Bohm period dependence on front-gate voltage.[30] The reported flux and charge superperiods, $\Delta_\Phi = 5h/e$ and $\Delta_Q = 2e$, were deduced theoretically using several FQH island models.[34-37] On the other hand, these periods were reported as either "not understood" in a Coulomb blockade model,[38] or even claimed as not possible in a composite fermion model[39] of the island.

Here we report experimental results obtained in a similar Fabry-Perot electron interferometer device, but with much larger 2D electron island, see inset in Fig. 1. The integer quantum Hall regime is used to determine the interferometer island area. In the FQH regime, the interfering $e/3$ quasiparticles execute a closed path around the island of the 2/5 FQH fluid containing $e/5$ quasiparticles. The 2D electron depletion, which largely determines the width of the $f = 1/3$ edge ring, does not depend on the device diameter. On the other hand, the enclosed 2/5 island is several times larger than before.[13,29-31] Hence, in this device, most of the island area is occupied by the 2/5 FQH fluid under coherent tunneling conditions, so that the directly-measured magnetic field period well approximates the flux period. We confirm the previously reported flux and charge superperiods of $\Delta_\Phi = 5h/e$ and $\Delta_Q = 2e$, respectively, both corresponding to addition of ten $e/5$ quasiparticles to the area enclosed by the interference path. These results are consistent with the Berry phase quantization condition that includes both Aharonov-Bohm and anyonic statistical contributions.[35]

## II. EXPERIMENTAL RESULTS

The electron interferometer device was fabricated from a low disorder AlGaAs/GaAs heterojunction crystal with 2D electrons ~320 nm below the surface.[40] The four independently-contacted front gates (FG) were defined by electron beam lithography on a pre-etched mesa with Ohmic contacts. After a shallow ~160 nm wet etching, Au/Ti front-gate metal was deposited in the etch trenches, followed by lift-off, inset in Fig. 1. The etch trenches define two ~1.1 µm lithographic width constrictions, which separate an approximately circular electron island from the 2D "bulk". Moderate front-gate voltages $V_{FG}$ are used to fine tune the constrictions for symmetry of the tunnel coupling and to increase the oscillatory interference signal. The shape of the electron density profile is predominantly determined by the etch trench depletion. The depletion potential has saddle points in the constrictions, and so has the resulting density profile. For the 2D bulk density $n_B = 1.0 \times 10^{11}$ cm$^{-2}$ there are ~4,500 electrons in the island.

The lithographic layout and dimensions of the present device are very similar to the device in Refs. 14 and 40, that has the entire island at filling 1/3 in the fractional regime. The two significant differences are: (i) the constriction-defining lip of the front gates is widened, and (ii) the etch trench depth is greater by ~20 nm. These relatively small differences combine to yield about three times more depleted constrictions, with the saddle point electron density estimated as ~0.78 of the island center density. This results in formation of a filling 1/3 edge ring passing through the constrictions, when the island and the 2D bulk both have FQH filling 2/5.

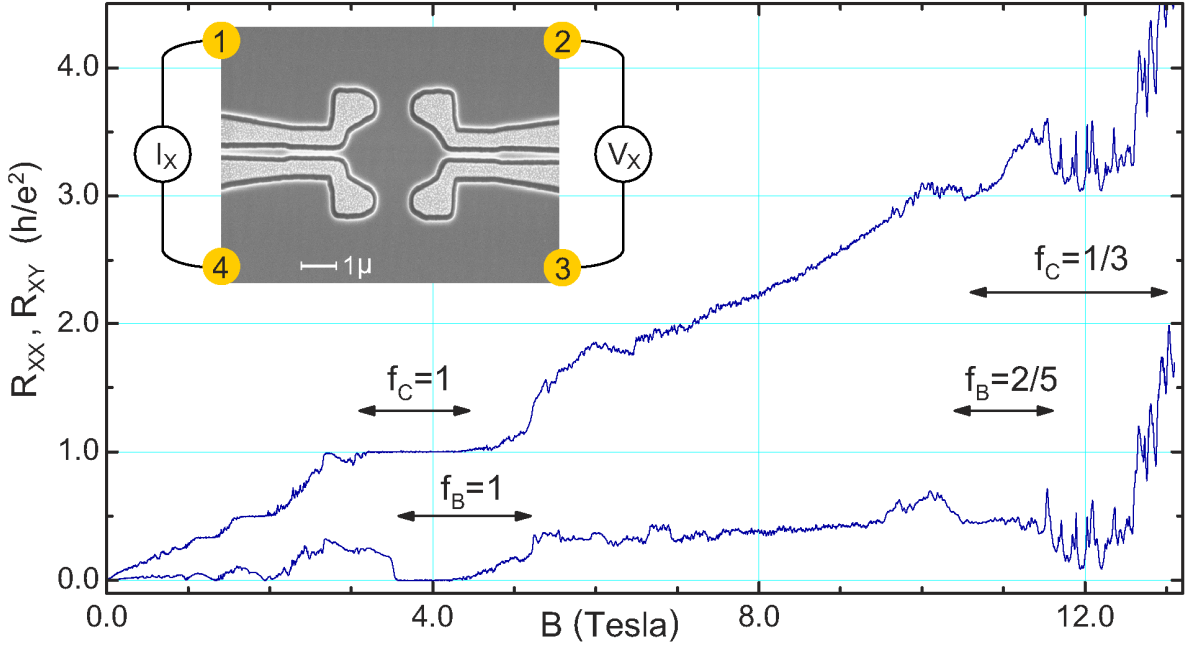

FIG. 1. The longitudinal $R_{XX}$ (lower trace) and Hall $R_{XY}$ magnetoresistance of the interferometer. The quantized plateaus (bulk $f_B$, constriction $f_C$) allow to determine the filling factor in the constrictions. The fine structure is due to quantum interference effects, sharp peaks are due to impurity-assisted tunneling. Inset: electron micrograph of the interferometer device. The front gates (light) are deposited in shallow etch trenches (dark). Depletion potential of the trenches defines the electron island. The edge channels circling the island are coupled by tunneling in the two constrictions, thus forming a Fabry-Perot interferometer. The backgate (not shown) extends over the entire 4×4 mm sample.

Samples were mounted on sapphire substrates with In metal, which serves as the global backgate, and were cooled in the tail of the mixing chamber of a $^3$He-$^4$He dilution refrigerator, immersed in the mixture. All data reported here were taken at 10.3 mK bath temperature, calibrated by nuclear orientation thermometry. The electromagnetic environment incident on the sample is attenuated by a combination of RF-lossy manganine wire ribbons and a series of cold low-pass RC network filters with a combined cut-off frequency ~50 Hz. Extensive cold filtering cuts the electromagnetic "noise" environment incident on the sample to ~7×10$^{-16}$ W, allowing to achieve an effective electron temperature $\leq 15$ mK in an interferometer device.[31]

Figure 1 shows longitudinal and Hall resistances in the interferometer sample with $V_{FG} \approx 60$ mV, similar to front-gate voltage in the oscillatory regime. Four-terminal resistance $R_{XX} = V_X / I_X$ was measured with 100 pA ($f = 1/3$) or 200 pA ($f = 1$), 5.4 Hz AC current injected at contacts 1 and 4. The resulting voltage $V_X$, including the Aharonov-Bohm oscillatory signal, was detected at contacts 2 and 3. The Hall resistance $R_{XY} = V_{4-2} / I_{3-1}$ is determined by the quantum Hall filling $f_C$ in the constrictions, giving definitive values of $f_C$. The oscillatory $\delta R$ is obtained from the directly measured $R_{XX}$ or $R_{XY}$ data after subtracting a smooth background. The conductance $\delta G$ is calculated from $\delta R$ and the quantized Hall resistance $R_{XY} = h / fe^2$ as $\delta G = \delta R / (R_{XY}^2 - \delta R R_{XY})$, a good approximation for $\delta R \ll R_{XY}$.

In the range of $B$ where the interference oscillations are observed, the counterpropagating edge channels must pass near the saddle points, where tunneling may occur.[13,14] Thus, the filling of the edge channels is determined by the saddle point filling. This allows to determine the saddle point density from the $R_{XX}(B)$ and $R_{XY}(B)$ magnetotransport; a systematic study of quantum Hall transport and analysis were reported for a similar sample in Ref. 41. The local Landau level filling $\nu = hn/eB$ is proportional to the local electron density $n$; accordingly the constriction $\nu_C$ is lower than the bulk $\nu_B$ in a given $B$. While $\nu$ is a variable, the quantum Hall exact filling $f$ is a quantum number defined by the *quantized* Hall resistance as $f = h/e^2 R_{XY}$.

In this device, the island center density is estimated to be close to the bulk $n_B$ at $V_{FG} = 0$, the constriction - island center density difference is ~20%. Thus, the whole island can be on the same plateau for strong quantum Hall states with wide plateaus, such as $f = 1$ and $1/3$. For example, in Fig. 1, there is a range of $B$ when both $f_C = 1$ and $f_B = 1$, as seen for $3.6\ \text{T} < B < 4.2\ \text{T}$, and both are $f = 1/3$ for $B > 12\ \text{T}$. The second possibility is an overlap of two plateaus with different filling. For example, $f_C = 1$ and $f_B = 4/3$, resulting in a *quantized* value of $R_{XX} = (h/e^2)(1/f_C - 1/f_B) = h/4e^2$, is seen at $B \approx 3.2\ \text{T}$, and $f_C = 1/3$ and $f_B = 2/5$, resulting in $R_{XX} = h/2e^2$, in the range $11.0\ \text{T} < B < 11.6\ \text{T}$ in Fig. 1. However, $f_C = 2$, $f_B = 3$ when $n_C \approx 0.67\ n_B$, e.g., is not seen in this sample.

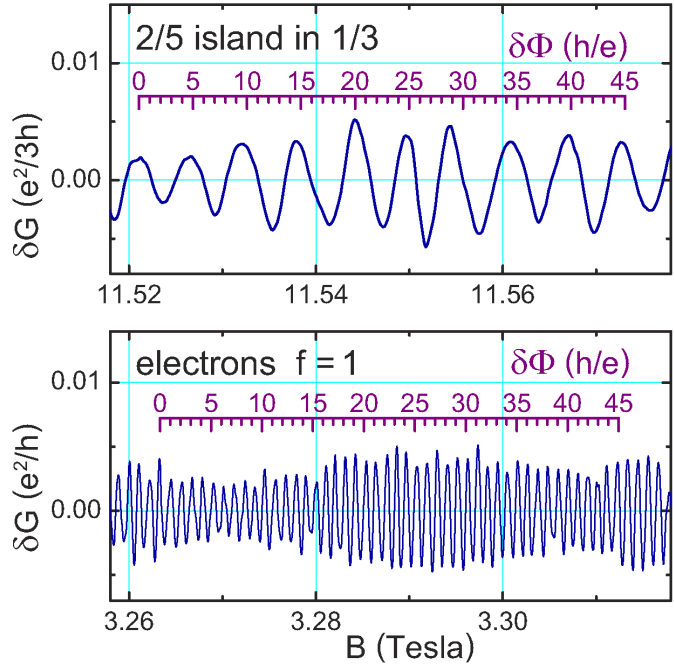

FIG. 2. Representative interference conductance oscillations for electrons, $f = 1$, and for $e/3$ quasiparticles in $f = 1/3$ edge channel circling around an island of 2/5 FQH fluid. Both are plotted on the same magnetic field scale, the magnetic field period ratio is 5.4±0.3. The flux scales are slightly different because the 2/5 island area is ~7% less than the $f = 1$ edge ring area.

In the integer quantum Hall regime the Aharonov-Bohm ring is formed by the two counter-propagating chiral edge channels passing through the constrictions.[20,21] Backscattering, which completes the interference path, occurs by quantum tunneling at the saddle points in the constrictions. The relevant particles are electrons of charge $-e$ and Fermi statistics, thus we can obtain an absolute calibration of the Aharonov-Bohm path area and the gate action of the interferometer. Fig. 2 shows conductance oscillations for $f = 1$; analogous oscillations for $f = 2$ were studied in this device, but are not reported here. The $f = 1$ magnetic field oscillation period is $\Delta_B = 1.06$ mT. The flux period here is $\Delta_\Phi = h/e$, this gives the interferometer path area $S = h/e\Delta_B = 3.91$ μm$^2$, the radius $r_{Out} = 1115$ nm.

We also observe the interferometric oscillations as a function of magnetic field in the FQH regime, when an $f = 1/3$ edge ring surrounds a 2/5 fluid island, Fig. 2. This occurs when the bulk 2/5 plateau and the constriction 1/3 plateau overlap, when the longitudinal $R_{XX} = (h/e^2)(3-5/2) \approx h/2e^2$. The magnetic field oscillation period in this regime is $\Delta_B = 5.7 \pm 0.3$ mT. Assuming the flux period is $\Delta_\Phi = 5h/e$, this gives the interferometer path area $S = 5h/e\Delta_B = 3.60$ μm$^2$, the radius $r_{In} = 1070$ nm. The conductance oscillations in this regime are found to be robust and reproducible, Fig. 3, systematically responding to a moderate change of front-gate voltage, as reported before for a smaller interferometer device.[30]

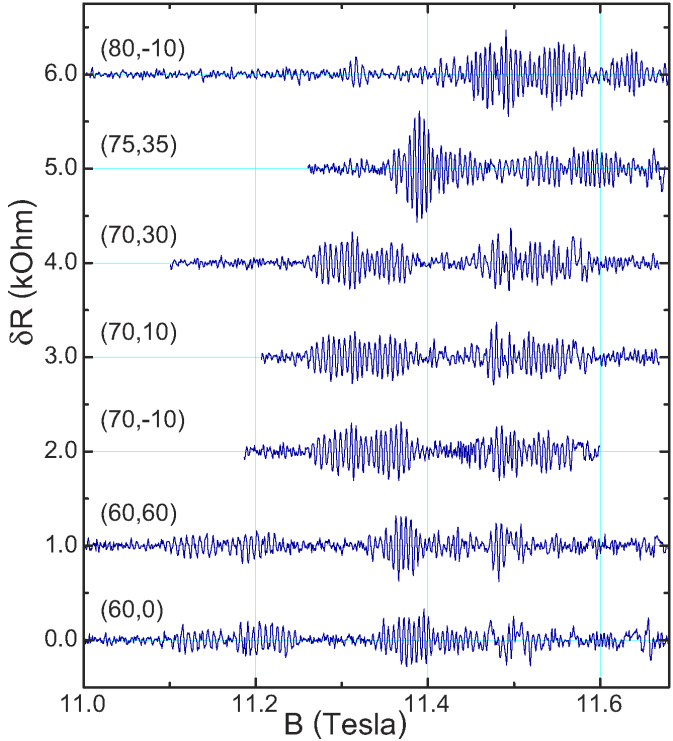

FIG. 3. Representative oscillatory $\delta R$ traces in the regime of $e/3$ quasiparticles encircling the 2/5 FQH island. Moderate front-gate $V_{FG}$ is applied, the $\delta R$ traces are labeled $(V_{FG1,2,3}, V_{FG4})$; the three voltages $V_{FG1,2,3}$ are equal. Successive traces are shifted by 1 kΩ. A positive front-gate voltage increases the island electron density and shifts the region of oscillations to higher $B$.

Classically, increasing $B$ by a factor of ~3 does not affect the electron density distribution in the island at all. Quantum corrections are expected to be small for a large island containing ~4,500 electrons.[42] Indeed, in experiments on a similar device, the $f = 1/3$ edge ring area was found to equal the integer value, within the ±3% experimental uncertainty.[14] As in the model of Ref. 13, in the fractional regime, the outer $f = 1/3$ edge ring of radius $r_{Out}$ encloses the 2/5 FQH island of radius $r_{In}$. The difference $r_{Out} - r_{In} \approx 45$ nm ($\approx 6\ell$, the magnetic length $\ell = \sqrt{\hbar/eB}$) approximates the width of the 1/3 incompressible ring. This width can be estimated from the model of Ref. 42: the incompressible edge "dipolar strip" width is $a_{1/3} = 50$ nm, where we use the value of the electron density gradient $[dn/dr]_{r=r_{Out}} = 3.6 \times 10^{20}$ m$^{-3}$ from a self-consistent island density model,[13,20] and the $f = 1/3$ FQH gap of 5 K at 12 T. The square of $a_{1/3}$ is proportional to gap and inversely proportional to the density gradient. Since the FQH gap is itself a weak function of $B$, $a_{1/3}$ is more sensitive to the gradient of the self-consistent island confining potential.

The ratio of the *magnetic field* periods $\Delta_B$ for the integer and fractional regime oscillations is 5.4 ± 0.3 in this sample. In interferometers with a smaller island (that also had somewhat different lithographic design), we reported the $\Delta_B$ ratio 7.15 for a $r_{Out} = 685$ nm, $r_{In} = 570$ nm device, and ratio 6.3±0.4 for a $r_{Out} = 920$ nm, $r_{In} = 820$ nm device.[13,20] Evidently, as the device area increases, the ratio of the magnetic field periods approaches 5 because the 2/5 FQH island occupies a larger part of the whole island area. Since the fundamental flux period is $h/e$ in the $f = 1$ integer regime, we conclude that the flux period is indeed $\Delta_\Phi = 5h/e$ when $e/3$ quasiparticles of the $f = 1/3$ FQH fluid execute a closed path around an island of the 2/5 fluid.

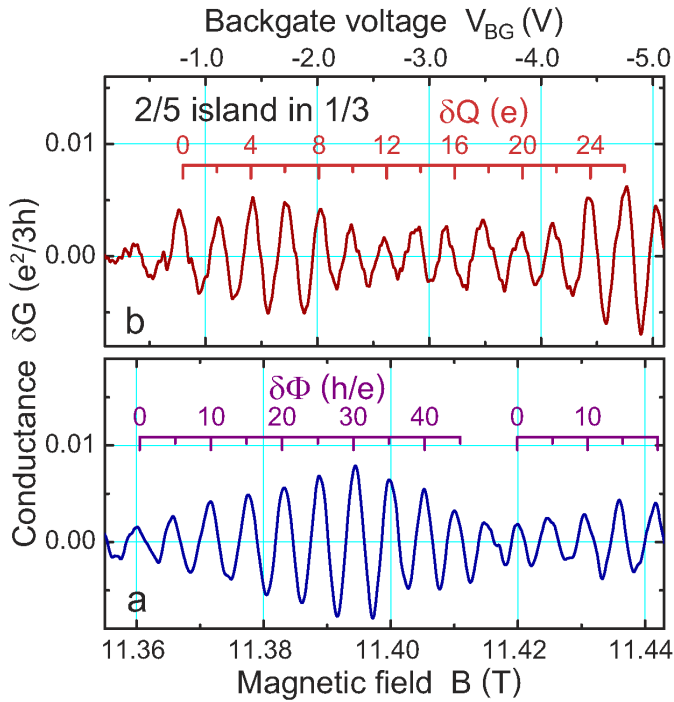

FIG. 4. A matched set of interference conductance oscillations in the regime of $e/3$ quasiparticles circling an island of the 2/5 FQH fluid. (a) Magnetic flux through the island period $\Delta_\Phi = 5h/e$ corresponds to creation of ten $e/5$ quasiparticles in the 2/5 fluid, two per $h/e$. (b) The backgate voltage island charging period $\Delta_Q = 2e = 10(e/5)$ agrees with incremental addition of ten $e/5$ quasiparticles. The ratio of the two periods confirms that the interference originates in the $f = 2/5$ FQH island. The interferometer device is calibrated using conductance oscillations for electrons, $f = 1$.

We use the backgate technique to measure the charge period in the fractional regime.[6,13,14,43] The backgate action $\delta Q / \delta V_{BG}$, where $Q$ is the electronic charge within the Aharonov-Bohm path, is calibrated with electrons in the integer regime. The calibration is done by evaluation of the coefficient $\alpha$ in $\Delta_Q = \alpha(\Delta_{V_{BG}} / \Delta_B)$, setting $\Delta_Q = e$ in the integer regime. Note that this procedure normalizes the backgate voltage periods by the experimental $B$-periods, canceling the variation in device area, for different devices and due to a front-gate bias. We could not calibrate $\alpha$ directly in the same device since there was a leakage present between the back- and front gates, that was observed to increase fast at lower magnetic fields. Instead, we use the coefficient $\alpha \approx 7.44e$ mT/V for the similar interferometer device fabricated from the same GaAs heterojunction wafer.[14]

Figure 4 shows the oscillations as a function of $V_{BG}$ in the fractional regime, and also the corresponding oscillations as a function of $B$. The front-gate voltage is the same for this matched set of complementary data. The periods are $\Delta_{V_{BG}} = 303$ mV and $\Delta_B = 5.61$ mT. Using the interferometer area obtained directly from the Aharonov-Bohm data, i.e., taking into account that $\Delta_B$ corresponds to five "flux quanta", we obtain $\Delta_Q = \alpha(5\Delta_{V_{BG}} / \Delta_B) = 2.01\,e$, equal (within the experimental uncertainty) to the expected value $\Delta_Q = 2e$.

In addition, the ratio $\Delta_{V_{BG}}/\Delta_B = 54.1$ V/T, multiplied by the calibration coefficient $\alpha/e$, is expected to give the ratio of electrons per "flux quanta", the quantum Hall filling $f$. Indeed, using the experimental periods we obtain $\alpha(\Delta_{V_{BG}}/e\Delta_B) = 0.403\pm0.01$, closely matching $f = 2/5$ and significantly distinct from $f = 1/3$. Thus, we conclude that the oscillations in Fig. 4 have the flux period $\Delta_\Phi = 5h/e$ and the charge period $\Delta_Q = 2e$, consistent with the prior report.[13] Using the $\Delta_{V_{BG}}/\Delta_B$ ratio technique and the matched (vs $V_{BG}$, vs $B$) data sets cancels, to first order, the dependence of the $V_{BG}$ and $B$ periods on the interferometer area and front-gate bias.

### III. ANALYSIS AND DISCUSSION

Experiments clearly show interference of Laughlin quasiparticles in an edge channel of the filling $f = 1/3$ FQH fluid, passing through the constrictions and circling an $f = 2/5$ island. Experimental tests establish: (i) the transport current displaying the interference signal is carried by the $e/3$ Laughlin quasiparticles, as evidenced by the Hall $R_{XY} = 3h/e^2$ and $R_{XY} = h/2e^2$, in Fig. 1 and in Fig. 4 in Ref. 13; (ii) the interference signal has magnetic flux period $\Delta_\Phi = 5h/e$ and the corresponding electric charge period $\Delta_Q = 2e$, see Figs. 2 and 4; (iii) these superperiods originate in an island that has the FQH filling 2/5, as is evident from the period ratio and is further supported by 2D electron island depletion modeling. These experimental superperiods do not violate gauge invariance,[33,44] and can be understood as follows.[35]

In an unbounded 2D FQH fluid, changing $\nu = hn/eB$ away from the exact filling $f$ is accomplished by creation of quasiparticles; the ground state consists of the $\nu = f$ condensate and the matching density of quasiparticles.[3-5,28] Starting at $\nu = f$, changing magnetic field adiabatically maintains the system in thermal equilibrium. The equilibrium electron density, determined by the positively charged donors, is not affected. In present geometry, changing $B$ also changes the flux $\Phi = BS$ through the semiclassical area $S$ enclosed by the interference path. At low temperature and excitation, the experiments probe the FQH ground state reconstruction within the interference path, in the large electron island, and the island is not isolated from the 2D bulk.

Thus, minimization of the total energy of the electron system by quasiparticle excitation in the large island is analogous to that in an unbounded 2D system. This holds as far as the Aharonov-Bohm oscillations are involved, which, in the ground state, are intimately connected with quasiparticle excitation. Changing filling $\nu$ by quasiparticle excitation eventually leads to a transition to the next FQH state. The island confining potential causes its edge state structure; this is also true in a large, but not infinite 2D electron system. As a transition from one quantum Hall ground state to another occurs, the edge channels move in space. Such effects are however related to transitions between neighboring quantum Hall states, change of Landau level filling $\nu$, not to the Aharonov-Bohm physics. Experimentally, periodic Aharonov-Bohm oscillations once in a while exhibit a jump, or a "phase slip". The phase slips [like that at $B \approx 11.417$ T in Fig. 4(a)] are presumably due to the secular edge channel movement related to changing $\nu$ that eventually causes the transition to the next quantum Hall plateau. The physics is different, however, and can be easily distinguished in a large device as not linked to the Aharonov-Bohm period. Note that $\nu$ does not depend on the device area, but Aharonov-Bohm period does. Thus, in a large area device, there are sequences of many periodic Aharonov-Bohm oscillations, occasionally interrupted by a "jump" due to edge channel movement on the microscopic scale.

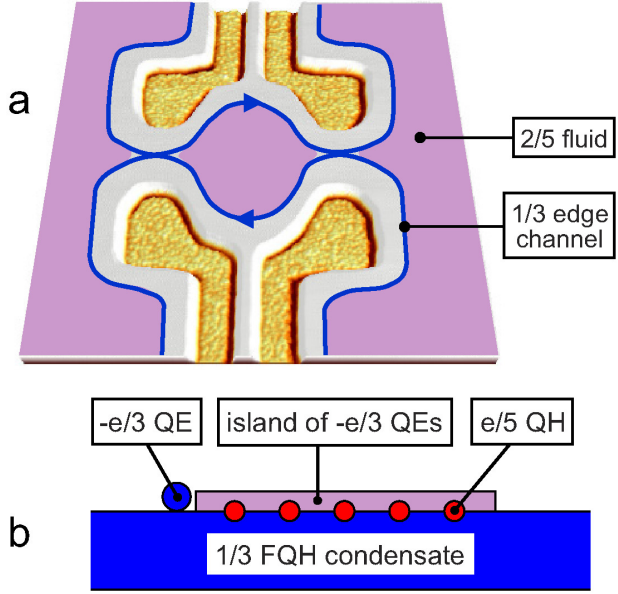

FIG. 5. (a) Atomic force (AFM) micrograph of the interferometer device with an illustration of the FQH filling profile. The transport current is carried in the 1/3 chiral edge channels. The path of the edge $-e/3$ quasielectrons is closed by tunneling in the two constrictions, and thus encircles the 2/5 island. (b) Illustration of the 2/5 island surrounded by 1/3 FQH fluid in the Haldane-Halperin hierarchy. The total 2D electron system is broken into three components: the incompressible exact filling 1/3 FQH condensate, the incompressible maximum density droplet of hierarchy $-e/3$ quasielectrons (QE), and the excited $e/5$ quasiholes (QH), appropriate for the $\nu < f = 2/5$ situation. A circling $-e/3$ QE is shown to the left of the island.

In the hierarchical construction,[28,12] the exact filling 2/5 FQH "daughter" condensate consists of a "maximum density droplet" of $-e/3$ quasielectrons in addition to the exact filling $f = 1/3$ condensate. The concentration of the $-e/3$ quasielectrons $n_{-e/3} = eB/5h$ is determined by their anyonic statistics. The resulting total *electron* charge density $en$ corresponds to the $f = 2/5$ exact filling condensate. Thus, the $f = 2/5$ island embedded in $f = 1/3$ FQH fluid can be understood as the island of $-e/3$ hierarchy quasielectrons on top of the $f = 1/3$ condensate, the 1/3 condensate extends beyond the quasielectron droplet and completely surrounds it, see Fig. 5.

The elementary charged excitations of the $f = 2/5$ condensate are the $\pm e/5$ quasielectrons and quasiholes, excited out of the condensate when the FQH fluid filling $\nu$ deviates from the exact filling 2/5. The density of the $\pm e/5$ quasiparticles can be obtained from conservation of the total electronic charge: $n_{\pm e/5} = \pm 5(f - \nu)eB/h$, where quasiholes are excited for $\nu < f$ and quasielectrons for $\nu > f$. In the island geometry, deviation of $\nu$ from $f$ also causes change in the number of the $-e/3$ hierarchy quasielectrons: $N_{-e/3} = n_{-e/3}S = SeB/5h$ in the island of area $S$. The two experimental methods of varying filling $\nu$ are: (i) sweeping the magnetic field $B$, and (ii) changing electron density $n$ by sweeping the backgate voltage at a fixed $B$. In experiments, either $B$ or $n$ vary very slowly, so that near thermal equilibrium is maintained at any time.

When $B$ is varied by a small $\delta B$, the equilibrium electron density profile (determined by the fixed positive background) is not affected except when transition to the next FQH state is considered, as discussed above. The island area is fixed by the large Coulomb energy, and the flux through the island $\Phi = BS$ is changed by $S\delta B$. The number of the $-e/3$ hierarchy quasielectrons in area $S$ is incremented by $Se\delta B/5h$. Concurrently, the $f = 2/5$ island condensate electron density changes by $fe\delta B/h$, which results in excitation of $e/5$ island quasiparticles, so as to maintain local charge neutrality of the total 2D electron system. Therefore, the *minimal* microscopic reconstruction of the island, the period $\delta B = \Delta_B$, occurs when one $-e/3$ hierarchy quasielectron is added, $Se\Delta_B/5h = 1$. This is exactly the observed $\Delta_\Phi = 5h/e$ flux periodicity. Within the period, increasing $B$, one $-e/3$ quasielectron is added to the island, the $f = 1/3$ condensate charge in area $S$ increases by $-5e/3$, and ten $+e/5$ island quasiholes are excited. The total island electronic charge remains the same, $-e/3 - 5e/3 + 10(e/5) = 0$, within the unchanged area $S$.

This process can be expressed in terms of the Berry phase $\gamma$ of the encircling $-e/3$ quasielectron, which includes the Aharonov-Bohm and the statistical contributions.[11,35] Ref. 11 used the adiabatic theorem to calculate the Berry phase of quasiholes in the $f = 1/3$ Laughlin wave function on a disc. When a quasihole adiabatically executes a closed path, the wave function acquires a Berry phase. Taking counterclockwise as the positive direction, they found the difference between an "empty" loop, containing the FQH condensate "vacuum" only, and a loop containing another quasihole to be $\Delta \gamma_{1/3} = 4\pi/3$, identified as the statistical contribution.

We define the statistics parameter of the particles $\Theta$ so that upon exchange the wave function acquires a phase factor $\exp(i\pi\Theta)$. Then $\Theta_{1/3} = \Theta_{-1/3}$ is the statistics of $\pm e/3$ quasiparticles of the $f = 1/3$ FQH fluid, and $\Theta_{2/5}^{-1/3}$ is involved when a $-e/3$ quasielectron encircles a $e/5$ quasihole of the $f = 2/5$ island fluid, the "mutual statistics" of different kinds of quasiparticles.[45,46] Ref. 35 derives and solves the Berry phase $\gamma$ equation describing the present experimental situation. It obtains Berry phase period $\Delta_\gamma = 2\pi$:

$$\frac{\Delta_\gamma}{2\pi} = -\frac{5}{3} + \Theta_{1/3} + 10\,\Theta_{2/5}^{-1/3} = 1. \tag{1}$$

Two concurrent physical processes comprise the period: increase by one in the number of island hierarchy $-e/3$ quasielectrons, and the excitation of ten $e/5$ quasiholes in the island. Thus, the physics under consideration leads to interpretation of Eq. (1) as two simultaneous equations, each with an integer Berry phase period:

$$1/3 + \Theta_{1/3} = 1, \text{ and} \tag{2a}$$

$$10\,\Theta_{2/5}^{-1/3} = 2. \tag{2b}$$

Eq. (2a) is identical to that obtained when only $e/3$ quasiparticles are present (no 2/5 island).[11,14,18] Eq. (2b) can be understood as sum of two $5\Theta_{2/5}^{-1/3} = 1$ equations, one for each of the two kinds of $e/5$ quasiparticles of the $f = 2/5$ condensate (the quantum numbers of the two kinds are expected to be identical). These equations are solved by $\Theta_{1/3} = 2/3$ and $\Theta_{2/5}^{-1/3} = 1/5$. The value $\Theta_{1/3} = 2/3$ is in agreement with the expectation and with recent experiments.[6,14,18] The value $\Theta_{2/5}^{-1/3} = 1/5$ appears to be consistent with what would be obtained in a Berry phase calculation similar to that of Ref. 11, by the Cauchy's theorem, including the charge deficiency in the 2/5 condensate created by excitation of an $e/5$ quasihole vortex, and maintaining the path of the adiabatically encircling $-e/3$ quasielectron fixed. Also, note that a $2.5h/e$ period (excitation of five island quasiparticles) were possible if $\Theta_{1/3}$ were an integer, that is, if the encircling $e/3$ quasiparticles were either bosons or fermions. Thus, the observed $5h/e$ superperiod requires both $\Theta_{2/5}^{-1/3}$ and $\Theta_{1/3}$ are anyonic. The relative (mutual) statistics of quasiparticles of the two FQH condensates at different filling are meaningful because both quasiparticle kinds are different collective excitations of a single highly correlated electron system comprising the parent-daughter FQH fluid with different fillings.

The same Berry phase equation describes the physically different process of the island charging by the backgate.[35,36] Here, in a fixed $B$, increasing positive $V_{BG}$ increases the 2D electron density.

The exact filling FQH condensate electron (and charge) density is fixed by the fixed $B$. The period consists of creating ten $-e/5$ quasielectrons out of the 2/5 FQH condensate within the interference path, while the path area increases by $5h/eB$ in the fixed $B$. Excitation of quasiparticles while the condensate density is fixed is possible because the condensate is not isolated from the bulk 2D electron system, and the charge imbalance is ultimately supplied from the contacts. Note that there is one more $-e/3$ hierarchy quasielectron in the 2/5 condensate of increased area $S + 5h/eB$. Thus, increasing $\nu$ by charging the island by the uniform electric field of the remote backgate is accommodated by creation of $-e/5$ quasielectrons and by concurrent outward shift of the 1/3 - 2/5 boundary, that is, the interference path. Ten $-e/5$ quasielectrons are excited out of the condensate (or, equivalently, ten quasiholes are absorbed into the condensate), the fixed condensate density is restored from the contacts, in constant $B$, the total FQH fluid electronic charge (condensate plus quasiparticles) changes by $-2e$ per $S$, the charge period.

Single-particle theory predicts Aharonov-Bohm flux period $\Delta_\Phi = 2\pi\hbar/q$ for charge $q$ particles.[32,33] This period is also expected for many-particle systems if the particle exchange statistics is integer, Fermi or Bose. In interacting many-electron systems, effective low-energy quasiparticles may have charge $q \neq e$. In the multiply-connected many-electron system, if a "fluxon" $h/e$ is added in the region of space from which the electrons are excluded (electron vacuum), the added flux can be annulled by a singular gauge transformation, leaving the many-electron system in the same state as before, and superperiods $> h/e$ are not possible even when $q < e$.[33,44] In our experiments, however, a uniform magnetic field is varied, rather than flux is inserted in the region of electron vacuum, and the situation is more subtle. The added flux results from increase in the applied magnetic field. The interacting electron system does reconstruct periodically, quasiparticles are excited, the many-electron system is not in the same microscopic state as before. Thus, gauge invariance does not preclude superperiods in the Fabry-Perot interferometer geometry, where there is no electron vacuum within the interference path.

## IV. CONCLUSIONS

In Section II we reported experiments on a large electron Fabry-Perot interferometer, where $e/3$ Laughlin quasiparticles execute a closed path around an island of the 2/5 FQH fluid. Most of the island area is occupied by the 2/5 FQH fluid, so that the directly-measured magnetic field period well approximates the flux period. The central experimental results obtained, that is, the flux and charge superperiods of $\Delta_\Phi = 5h/e$ and $\Delta_Q = 2e$, are robust and do not involve any adjustable parameter fitting to a model. In Section III we presented a microscopic model of the origin of the superperiod based on the Haldane-Halperin fractional-statistics hierarchical construction of the 2/5 FQH fluid. The superperiod comprises incrementing by one the state number of the $-e/3$ quasielectron circling the island and concurrent excitation of ten $e/5$ quasiparticles in the island 2/5 fluid. Variation of the magnetic field does not affect the charge state of the island. Quantization of the Berry phase of the circling $e/3$ quasiparticles in integer multiples of $2\pi$ gives anyonic statistics $\Theta_{1/3} = 2/3$ for the $e/3$ quasiparticles, and $\Theta_{2/5}^{-1/3} = 1/5$, the mutual statistics, when a $-e/3$ quasielectron encircles a $e/5$ quasihole of the 2/5 fluid.

### ACKNOWLEDGMENTS

Stimulating discussions with D. V. Averin, P. Bonderson, D. E. Feldman, E. Fradkin, B. I. Halperin, T. H. Hanson, S. A. Kivelson, J. M. Leinaas, C. Nayak, K. Shtengel, S. H. Simon, A. Stern, and F. Wilczek are gratefully acknowledged. This work was supported in part by the National Science Foundation under grant DMR-0555238.

1. K. von Klitzing, G. Dorda, and M. Pepper, Phys. Rev. Lett. **45**, 494 (1980).
2. D. C. Tsui, H. L. Stormer, and A. C. Gossard, Phys. Rev. Lett. **48**, 1559 (1982).
3. R. B. Laughlin, Phys. Rev. Lett. **50**, 1395 (1983).
4. *The Quantum Hall Effect*, 2nd Ed., edited by R. E. Prange and S. M. Girvin (Springer, NY, 1990).
5. D. Yoshioka, *The Quantum Hall Effect* (Springer, NY, 2002).
6. V. J. Goldman and B. Su, Science **267**, 1010 (1995).
7. L. Saminadayar, D. C. Glattli, Y. Jin, and B. Etienne, Phys. Rev. Lett. **79**, 2526 (1997).
8. R. De-Picciotto *et al.*, Nature (London) **389**, 162 (1997).
9. J. M. Leinaas and J. Myrheim, Nuovo Cimento B **37**, 1 (1977).
10. F. Wilczek, Phys. Rev. Lett. **48**, 1144 (1982).
11. D. Arovas, J. R. Schrieffer, and F. Wilczek, Phys. Rev. Lett. **53**, 722 (1984).
12. B. I. Halperin, Phys. Rev. Lett. **52**, 1583 (1984).
13. F. E. Camino, W. Zhou, and V. J. Goldman, Phys. Rev. B **72**, 075342 (2005).
14. F. E. Camino, W. Zhou, and V. J. Goldman, Phys. Rev. Lett. **98**, 076805 (2007).
15. F. Wilczek, arXiv:0808.2448 (2008).
16. C. Nayak, S. H. Simon, A. Stern, M. Freedman, and S. Das Sarma, Rev. Mod. Phys. **80**, 1083 (2008).
17. P. Bonderson, A. Kitaev, and K. Shtengel, Phys. Rev. Lett. **96**, 016803 (2006).
18. V. J. Goldman, J. Liu, and A. Zaslavsky, Phys. Rev. B **71**, 153303 (2005).
19. B. J. van Wees *et al.*, Phys. Rev. Lett. **62**, 2523 (1989).
20. F. E. Camino, W. Zhou, and V. J. Goldman, Phys. Rev. B **72**, 155313 (2005).
21. F. E. Camino, W. Zhou, and V. J. Goldman, Phys. Rev. B **76**, 155305 (2007).
22. M. D. Godfrey *et al.*, arXiv:0708.2448 (2007).
23. Y. Zhang *et al.*, Phys. Rev. B **79**, 241304 (2009).
24. R. L. Willett, L. N. Pfeiffer, and K. W. West, PNAS **106**, 8853 (2009).
25. P. V. Lin, F. E. Camino, and V. J. Goldman, Phys. Rev. B **80**, in press; arXiv:0902.0811v.2.
26. D. T. McClure *et al.*, arXiv:0903.5097 (2009).
27. C. de C. Chamon, D. E. Freed, S. A. Kivelson, S. L. Sondhi, and X. G. Wen, Phys. Rev. B **55**, 2331 (1997).
28. F. D. M. Haldane, Phys. Rev. Lett. **51**, 605 (1983).
29. F. E. Camino, W. Zhou, and V. J. Goldman, Phys. Rev. Lett. **95**, 246802 (2005).
30. W. Zhou, F. E. Camino, and V. J. Goldman, Phys. Rev. B **73**, 245322 (2006).
31. F. E. Camino, W. Zhou, and V. J. Goldman, Phys. Rev. B **74**, 115301 (2006).
32. Y. Aharonov and D. Bohm, Phys. Rev. **115**, 485 (1959).
33. Y. Aharonov and D. Bohm, Phys. Rev. **123**, 1511 (1961).
34. E. A. Kim, Phys. Rev. Lett. **97**, 216404 (2006).
35. V. J. Goldman, Phys. Rev. B **75**, 045334 (2007).
36. G. A. Fiete, G. Refael, and M. P. A. Fisher, Phys. Rev. Lett. **99**, 166805 (2007).
37. Z. X. Hu, H. Chen, K. Yang, E. H. Rezayi, and X. Wan, Phys. Rev. B **78**, 235315 (2008).
38. B. Rosenow and B. I. Halperin, Phys. Rev. Lett. **98**, 106801 (2007).
39. J. K. Jain and C. Shi, Phys. Rev. Lett. **96**, 136802 (2006).
40. M. Shayegan *et al.*, Appl. Phys. Lett. **53**, 2080 (1988).
41. P. V. Lin, F. E. Camino, and V. J. Goldman, Phys. Rev. B **78**, 245322 (2008).
42. D. B. Chklovskii, B. I. Shklovskii, and L. I. Glazman, Phys. Rev. B **46**, 4026 (1992).
43. V. J. Goldman, I. Karakurt, J. Liu, and A. Zaslavsky, Phys. Rev. B **64**, 085319 (2001).
44. N. Byers and C. N. Yang, Phys. Rev. Lett. **7**, 46 (1961).
45. F. Wilczek, Phys. Rev. Lett. **69**, 132 (1992).
46. W. P. Su, Y. S. Wu, and J. Yang, Phys. Rev. Lett. 77, 3423 (1996).